\documentclass[twocolumn, tighten]{aastex631}

\shorttitle{Is Cloud-9 a RELHIC?}
\shortauthors{Benitez-Llambay}

\defcitealias{Zhou2023}{Z23}
\defcitealias{Benitez-Llambay2020}{BLF20}
\defcitealias{Benitez-Llambay2017}{BL17}

\begin{document}

\title{Is a recently discovered HI cloud near M94 a starless dark matter halo?}

\correspondingauthor{Alejandro Benitez-Llambay}
\email{alejandro.benitezllambay@unimib.it}

\author[0000-0001-8261-2796]{Alejandro Benitez-Llambay}
\affil{Dipartimento di Fisica G. Occhialini, Universit\`a degli Studi di Milano Bicocca, Piazza della Scienza, 3 I-20126 Milano MI, Italy}
\author[0000-0003-3862-5076]{Julio F. Navarro}
\affil{Department of Physics and Astronomy, University of Victoria, Victoria, BC V8P 5C2, Canada}

\begin{abstract}

Observations with the Five-Hundred-Meter Aperture Spherical Telescope have revealed the presence of a marginally-resolved source of 21 cm emission from a location $\sim50'$ from the M94 galaxy, without a stellar counterpart down to the surface brightness limit of the DESI Imaging Legacy Survey ($\sim29.15$ mag arcsec$^{-2}$ in the $g$ band). The system (hereafter Cloud-9) has round column density isocontours and a line width consistent with thermal broadening from gas at $T\sim2\times10^4$ $K$. These properties are unlike those of previously detected dark HI clouds and similar to the expected properties of REionization-Limited-HI Cloud (RELHICs), namely, starless dark matter (DM) halos filled with gas in hydrostatic equilibrium and in thermal equilibrium with the cosmic ultraviolet background. At the distance of M94, $d\sim4.7$ Mpc, we find that Cloud-9 is consistent with being a RELHIC inhabiting a Navarro-Frenk-White (NFW) DM halo of mass, $M_{200}\sim5\times10^{9}$ $M_{\odot}$, and concentration, $c_{\rm NFW}\sim13$. Although the agreement between the model and observations is good, Cloud-9 appears to be slightly, but systematically, more extended than expected for $\Lambda$CDM RELHICs. This may imply either that Cloud-9 is much closer than implied by its recessional velocity, $v_{\rm CL9}\sim300$ km s$^{-1}$, or that its halo density profile is flatter than NFW, with a DM mass deficit greater than a factor of $10$ at radii $r\lesssim1$ kpc. Further observations may aid in constraining these scenarios better and help elucidate whether Cloud-9 is the first ever observed RELHIC, a cornerstone prediction of the $\Lambda$CDM model on the smallest scales.
\end{abstract}

\keywords{Dark matter(353) --- Cosmology(343) --- Reionization(1383) --- }

\section{Introduction} \label{sec:intro}

A distinctive prediction of the Lambda-Cold Dark Matter ($\Lambda$CDM) model of structure formation is the existence of a vast number of collapsed halos, whose density follows a universal profile~\citep[][hereafter NFW]{Navarro1996}, and whose abundance at the low-mass end scales as a power law of the mass, $\propto M^{-1.9}$~\cite[e.g.][]{Press1974, Bond1991, Jenkins2001, Angulo2012, Wang2020}. This result, combined with the relatively flat faint-end of the galaxy luminosity function, implies that a large population of low-mass dark matter (DM) halos must remain ``dark'' or starless until the present day~\cite[see, e.g.,][and references therein]{Ferrero2012}.

The origin of these ``dark'' halos in $\Lambda$CDM is well motivated theoretically: galaxies can only form in the center of halos whose mass exceeds a redshift-dependent critical mass, $M_{\rm crit}(z)$. This critical mass corresponds, before cosmic reionization, to the halo mass above which atomic cooling becomes efficient~\citep[e.g.][]{Blumenthal1984, Bromm2011}, and after reionization, to the halo mass above which the pressure of the photoheated gas cannot overcome the gravitational force of the halo~\citep[][hereafter BLF20]{Benitez-Llambay2020}. 

Analytical models~\citep[][]{Ikeuchi1986, Rees1986, Benitez-Llambay2020} and results from hydrodynamical simulations~\citep[e.g.][]{Hoeft2006, Okamoto2008, Benitez-Llambay2017}, demonstrate that DM halos less massive than $M_{\rm crit}\sim7\times10^{9}$ $M_{\odot}$ today should contain gas in hydrostatic equilibrium with the gravitational potential of the halo and in thermal equilibrium with the external ultraviolet background radiation (UVB). Moreover, the models indicate that halos that never exceeded $M_{\rm crit}(z)$ should remain devoid of stars to the present day. 

For the most massive ``dark'' halos, the high density and low temperature of their gas lead to the formation of neutral hydrogen (HI) in the center, making them detectable in 21 cm emission. This is why these systems were termed ``Reionization-Limited-HI-Clouds'' (RELHICs) by~\citet[][]{Benitez-Llambay2017} (hereafter, BL17), and are analogues of the mini-halos envisaged by~\cite{Rees1986} and~\cite{Ikeuchi1986} in the context of the early Ly$\alpha$ forest models.

The properties of RELHICs in $\Lambda$CDM were studied by~\citetalias{Benitez-Llambay2017}. These authors concluded that RELHICs should be nearly spherical extragalactic gas clouds in hydrostatic equilibrium with the underlying NFW halo. Their gas density profile is well specified because of the distinctive density-temperature relation that arises from the interplay between gas cooling and photoheating. As RELHICs are close to hydrostatic equilibrium, they lack significant velocity dispersion.

Detecting RELHICs would represent a remarkable achievement mainly for two reasons. Above all, it would unequivocally confirm the presence of bound collapsed DM structures on mass scales below galaxies —a pivotal prediction of the $\Lambda$CDM model. Secondly, it would pave the way towards a novel and independent way to probe $\Lambda$CDM on small scales, where $\Lambda$CDM is still subject to heavy scrutiny~\citep[see, e.g.,][for a recent review of the small-scale challenges faced by $\Lambda$CDM]{Bullock2017}.

As discussed by~\citetalias{Benitez-Llambay2017}, the most promising RELHIC candidates to date have been some of the Ultra Compact High-Velocity Clouds (UCHVCs) identified in the ALFALFA catalog~\citep{Adams2013, Haynes2018}. This catalog contains roughly 60 ``dark'' HI clouds whose sizes and fluxes are broadly consistent with RELHICs. Of these candidates, the systems that appear round in the sky display either a large broadening of their HI line compatible with non-zero velocity dispersion (or rotation) or negative recessional velocity, indicating they are likely nearby sources. On the other hand, HI clouds receding from us and having a small line width broadening display a highly irregular morphology. Thus, no observational analog entirely consistent with RELHICs has been positively identified to date.

In this work, we focus on the discovery by~\cite{Zhou2023} (hereafter Z23) of extended emission in an isolated field near M94. The system (termed Cloud-9; hereafter CL-9) was observed with the Five-Hundred-Meter Aperture Spherical Telescope (FAST), has no obvious luminous counterpart, and exhibits properties consistent with those expected for RELHICs. Using the models introduced by~\citetalias{Benitez-Llambay2017} and~\citetalias{Benitez-Llambay2020}, we address whether the~\citetalias{Zhou2023} observations are consistent with CL-9 being a $\Lambda$CDM RELHIC. We refer the interested reader to those papers for further details.

\begin{figure}
    \centering
    \includegraphics[]{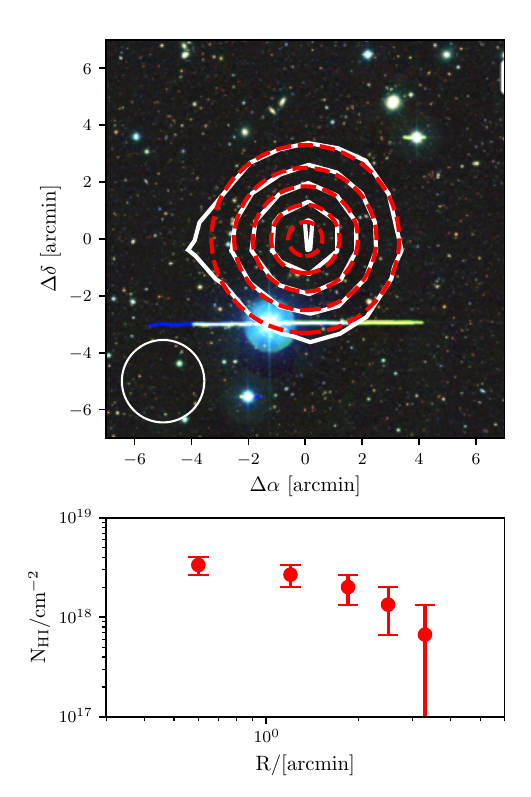}
    \caption{CL-9's observed column density isocontours~\citep[taken from][]{Zhou2023} superimposed on a DESI LS color image of the same field. The white circle indicates the FAST beam size. We approximate the observed isocontours with the dashed circles to construct the observed column density profile displayed in the bottom panel. The error bars indicate $3\sigma$ uncertainties. The coordinates are relative to the origin, $(\alpha, \delta)=(12^h51^m52^s,+40^o17'29'')$.}
    \label{Fig:isocontours}
\end{figure}

\section{Method}

\subsection{Observations}
\label{Sec:observations}

Recently,~\citetalias{Zhou2023} reported the detection of CL-9, a relatively isolated HI cloud without a luminous counterpart brighter than the surface brightness limit of the DESI Legacy Imaging Survey (DESI LS), namely, 29.15, 28.73, and 27.52 mag arcsec$^{-2}$ for the $g,r,z$ filters, respectively~\citep{Martinez-Delgado2023}. The system is at a projected angular distance $\sim51.87'$ from the center of M94, a galaxy located at a distance, $d\sim4.66$ Mpc~\cite[e.g.][]{Lee2011}. This value is broadly consistent with the distance obtained from its radial velocity, $v_{\rm M94}\sim287$ km $s^{-1}$, assuming it is receding from us on the Hubble flow. Other distance estimates for M94 also place the galaxy in the distance range, $4\lesssim d$ / Mpc $\lesssim5$,~\citep[e.g.,][]{Karachentsev2004, Crook2007, Cappellari2011, Tully2016, Karachentsev2018}. 

CL-9 has a recessional velocity similar to that of M94, $v_{\rm CL9}\sim300$ km s$^{-1}$~\citepalias{Zhou2023}. This coincidence, together with the close angular separation, makes it likely that CL-9 is in the vicinity of M94. Assuming this is the case, the projected distance between CL-9 and M94 corresponds to a physical separation greater than $\sim70$ kpc and to a maximum stellar mass for CL-9, $M_{\rm str}\lesssim10^{5}$ $M_{\odot}$, as reported by~\citetalias{Zhou2023}. 

If CL-9 is not near M94, then its recessional velocity makes it unlikely that the system is closer than 3 Mpc from us. Indeed, no galaxies with a reliable distance estimate closer than 3 Mpc have recessional velocities that reach this value~\cite [see, e.g.,][]{Karachentsev2019}. 

This lower bound on the distance is further supported by the distance estimate based on the velocity field reconstruction of the Local Volume using CosmicFlows-3~\citep{Tully2016}, which returns a distance in the range $3\lesssim d$/Mpc $\lesssim4$. The lower/higher value is obtained when the reconstruction uses the Numerical Action Method~\citep[][]{Shaya2017}/Wiener filter model~\citep{Graziani2019}\footnote{We queried the distance using the CosmicFlows-3 calculator available at \url{http://edd.ifa.hawaii.edu}~\citep{Kourkchi2020}.}. However, it is not possible to exclude the possibility that CL-9 is farther than M94 using the system's recessional velocity alone.

CL-9 appears round in the sky and displays a narrow broadening of its emission line ($W_{50}\sim20$ km s$^{-1}$ at its peak column density), consistent with thermal broadening arising from gas at $T\sim2\times10^4$ $K$. These properties are consistent with those expected for RELHICs~\citepalias{Benitez-Llambay2017}. We note, however, that the inferred shape of CL-9 may be affected by the large FAST beam, whose size is comparable to the spatial extent of the detection. 

CL-9 is unlikely to be a self-gravitating system. If the system's sound-crossing time equals the free-fall time within the observed range, then the required HI mass for the system to be in equilibrium (given its linewidth and size at the distance of M94) is $M_{\rm HI}\sim4\times10^{8}$ $M_{\odot}$. This value is orders of magnitude higher than the derived HI mass given its total flux ($M_{\rm HI}\sim7\times10^5$ $M_{\odot}$)~\citepalias{Zhou2023}, implying the presence of a large amount of gravitational mass other than neutral hydrogen.

We show CL-9's observed column density isocontours, taken from the work of~\citetalias{Zhou2023}, in the top panel of Fig.~\ref{Fig:isocontours}. Following~\citetalias{Zhou2023}, we superimpose the contours over a DESI LS color image to emphasize that there is no obvious extended luminous counterpart within the surface brightness limit of the survey\footnote{Note that the presence of a bright star near CL-9 may affect somewhat the exact surface brightness limit reached by the DESI LS image at this location.}. The outermost isocontour corresponds to a value equal to the $3\sigma$ detection limit, $N_{\rm HI}=6.7\times10^{17}$ cm$^{-2}$, and the contour values increase in steps of $6.7\times10^{17}$ cm$^{-2}$, so that the maximum column density reached by the innermost contour is, $N_{\rm HI}=3.35\times10^{18}$ cm$^{-2}$. 

Because the system's isocontours are round, we approximate them by the circles depicted by the dashed lines and use the circles' radius to produce the column density profile shown by the red dots in the bottom panel of Fig.~\ref{Fig:isocontours}. Since the innermost isocontour is elliptical, we will not use it for our analysis. 

\subsection{RELHICs}
\subsubsection{Intrinsic column density profile and mock observations}

We model RELHICs following~\citetalias{Benitez-Llambay2017}. This implies assuming that RELHICs are spherical gaseous systems in hydrostatic equilibrium with an NFW DM halo and in thermal equilibrium with a~\cite{Haardt2012} UVB. We note that the presence of a stellar counterpart does not affect the structure nor the system's stability, provided the stars are negligible contributors to the gravitational potential. 

To solve the hydrostatic equilibrium equation, we use a boundary condition where the pressure at infinity equals the pressure of the intergalactic medium at the mean density of the Universe. With this condition, the model reproduces the detailed structure of stable gaseous halos in large high-resolution cosmological hydrodynamical simulations~\citepalias{Benitez-Llambay2017,Benitez-Llambay2020}. In this model, RELHICs are characterized by a distinctive maximum central density, which depends on the gas temperature, halo virial mass, $M_{200}$, and concentration, $c_{\rm NFW}$. 

To derive the HI density profile of RELHICs, we apply the ~\cite{Rahmati2013} results. Once the HI density profile is known, we calculate the intrinsic HI column density by projecting the HI density. 

To compare~\citetalias{Zhou2023} observations with a RELHIC model, we place RELHICs at the observed distance and convolve their intrinsic column density profile with a circular Gaussian beam with standard deviation, $\sigma_{\rm beam}=1.23'$, whose full width at half maximum matches that of the FAST beam, $\sim 2.9'$. Performing this convolution is crucial because we will compare models with observations on scales smaller than the FAST beam size. Moreover, at the M94 distance, the angular extent of the central HI core of RELHICs is comparable to the beam size. 

\section{Results}

\subsection{CL-9 as a $\Lambda$CDM RELHIC}

We now address whether the observed CL-9's column density profile is consistent with the system being a RELHIC. To this end, we consider RELHICs embedded within an NFW DM halo, a profile characterized by the halo concentration and virial mass. These two parameters fully specify the RELHICs' total HI mass, central density, and characteristic size.

\begin{figure}
    \centering
    \includegraphics[]{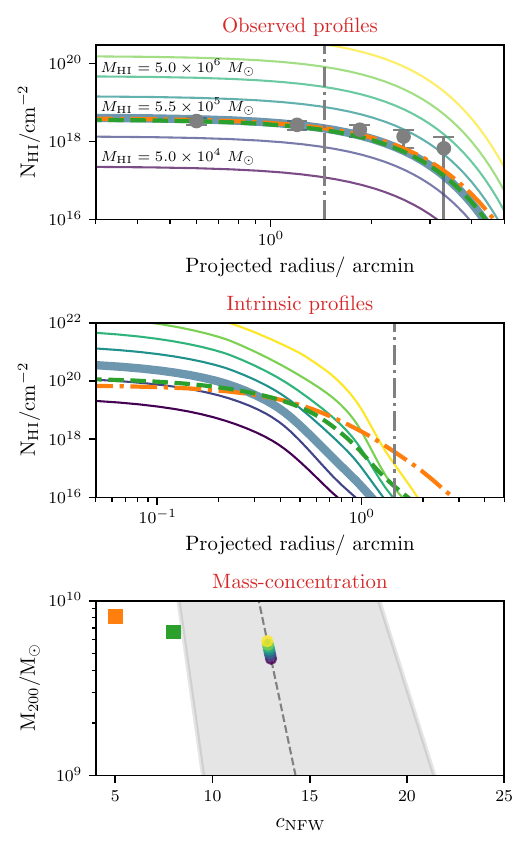}
    \caption{\textbf{Top panel}: CL-9's observed column density profile (circles), including $3\sigma$ uncertainties, together with mock-observed RELHICs (lines) at the distance of M94 and convolved with the FAST Gaussian beam. RELHICs were chosen to bracket CL-9's observations while following the $\Lambda$CDM mass-concentration relation. The thick line highlights the best-fit model (see text for discussion), and the labels indicate the total HI mass of the model immediately below\textbf{Middle panel}: intrinsic column density profiles of the models shown in the top panel. The vertical line indicates the radial extent of the FAST beam. \textbf{Bottom panel}: mass-concentration relation of the models, together with the Ludlow et al. (2016) mass-concentration relation (line) and scatter (shaded region). The orange and green squares indicate examples with lower-than-average concentrations that also match observations (see the lines of the same color in the middle and top panels).}
    \label{Fig:LCDM}
\end{figure}

Therefore, we construct a grid of models as a function of virial mass and concentration, imposing the \cite{Ludlow2016} mass-concentration relation. We place the models at a fiducial distance of M94, $d=4.66$ Mpc, and convolve them with the FAST Gaussian beam. We then adopt the best model as the model that matches CL-9's column density at the location of the second innermost isocontour, i.e., the highest signal-to-noise isocontour that has a reliable distance estimate from the center. 

We show the model results, compared with CL-9's observation, in the top panel of Fig.~\ref{Fig:LCDM}. The thin curves show examples of $\Lambda$CDM RELHICs within a very narrow range of halo mass centered at the mass of the best RELHIC model (thick line). 

Three outcomes of this exercise deserve particular attention. Firstly, it is remarkable that it is possible to match CL-9's isocontours by varying solely the halo mass of a RELHIC at the distance of M94 without further adjustments. Secondly, if CL-9 is indeed a RELHIC, its observed column density profile imposes a tight constraint on the mass of its DM halo, as small departures in halo mass relative to the best model produce models whose central column density quickly departs from observations. This implies that the derived properties are not very sensitive to the comparison between the model and observations at the adopted isocontour. Thirdly, the best RELHIC model contains an HI mass, $M_{\rm HI}\sim5.5\times10^5$ $M_{\odot}$, which
is in excellent agreement with the value derived by~\citetalias{Zhou2023} for CL-9\footnote{CL-9's integrated flux is $S_{21}=0.14\pm0.02$ Jy km s$^{-1}$~\citepalias{Zhou2023}, implying an HI mass, $M_{\rm HI}\sim(7.2\pm1)\times10^{5}$ $M_{\odot}$ at the M94 distance.}. In contrast, other models contain HI masses that significantly depart from the inferred value, as indicated by the labels in the top panel of Fig.~\ref{Fig:LCDM}. This demonstrates that CL-9 is fully consistent with a RELHIC even if it was considered an unresolved source.

We thus conclude that if CL-9 is at the distance of M94, then its total HI mass and column density profile properties are fully consistent with a $\Lambda$CDM RELHIC of mass $M_{200}\sim5.04\times10^{9}$ $M_{\odot}$, and concentration, $c_{\rm NFW}=12.94$. 

Although the best $\Lambda$CDM RELHIC shown by the thick solid line in Fig.~\ref{Fig:LCDM} is consistent with observations, there are slight but systematic differences between the best-fit model and observations at larger radii. A priori, these could originate from: 1) CL-9 inhabiting a DM halo with lower-than-average concentration; 2) a wrong distance estimate to CL-9; and 3) departures of the structure of the DM halo compared to $\Lambda$CDM expectations. 

To address the first possibility, we constructed models with a lower-than-average concentration that match the second innermost CL-9's isocontour. The orange and green squares in the bottom panel of Fig.~\ref{Fig:LCDM} show two extreme examples. Lowering the concentration increases the halo mass, but only mildly, demonstrating that the leading factor determining the central column density of RELHICs is the DM mass. In addition, the "observed" models are only marginally more extended than the fiducial best-fit $\Lambda$CDM RELHIC (see the dashed and dot-dashed lines in the top and middle panels), indicating that halo concentration cannot resolve the tension between modeling and observations if CL-9 is at the distance of M94. 

We explore the other two possible sources of discrepancies in the following sections.

\subsection{Is CL-9 a $\Lambda$CDM RELHIC closer than M94?}

If CL-9 is indeed a $\Lambda$CDM RELHIC, the systematic difference between the model and observations in the outer regions may simply reflect a wrong distance estimate for CL-9; RELHICs would appear more extended in projection and be more consistent with CL-9 if we place the system closer to the observer. 

\begin{figure}
    \centering
    \includegraphics[]{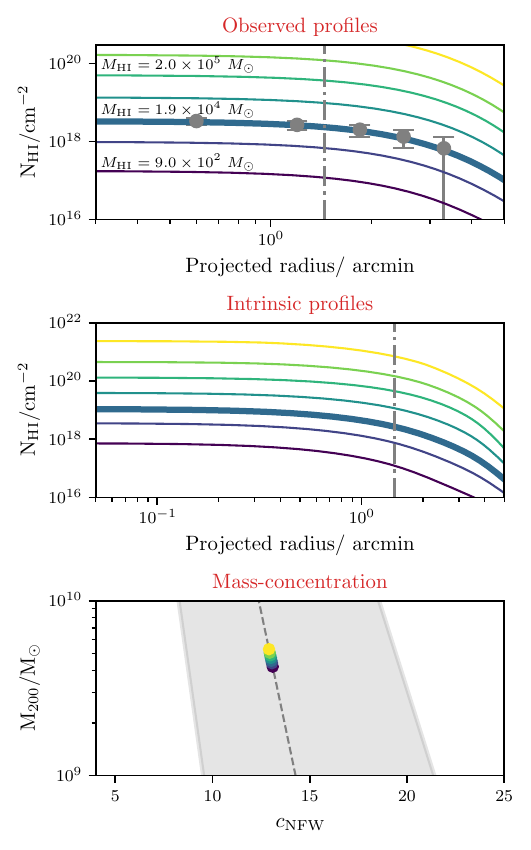}
    \caption{Identical to Fig.~\ref{Fig:LCDM}, but assuming CL-9 is at a distance, $d=500$ kpc.}
    \label{Fig:density_profiles_low_distance}
\end{figure}

To explore the possibility that CL-9 is much closer than M94, we varied the distance to CL-9 and found that bringing the system to a distance $d=500$ kpc from us improves the quality of the fit while still adopting the mean $\Lambda$CDM mass-concentration relation. This is shown in the top panel of Fig.~\ref{Fig:density_profiles_low_distance}, which is analogous to Fig.~\ref{Fig:LCDM} but assumes $d=500$ kpc.

Although this smaller distance improves the agreement between the model and observations, it is disfavored by CL-9's high recessional velocity, as discussed in Sec.~\ref{Sec:observations}. Placing CL-9 much farther would only increase its neutral hydrogen mass without improving the fit to observations. As shown by~\citetalias{Benitez-Llambay2017}, RELHICs should have an HI mass, $M_{\rm HI}\lesssim3\times10^{6}$ $M_{\odot}$, which places an upper limit to CL-9's distance, $d_{\rm CL9}\lesssim10$ Mpc, if the system is indeed a RELHIC.

Finally, changes in the distance estimate of CL-9 only have a minor impact on the derived DM halo parameters. Although we have changed CL-9's distance by almost an order of magnitude between Fig.~\ref{Fig:LCDM} and Fig.~\ref{Fig:density_profiles_low_distance}, the resulting DM mass of the best model remains similar between the two models. It is now: $M_{200}=4.5\times10^{9}$ $M_{\odot}$ ($c_{\rm NFW}=13.02$). This is because RELHICs' neutral hydrogen density is extremely sensitive to halo mass, thus making the distance a secondary parameter in the explored range. This is not the case for the HI mass, which depends on distance. The inferred HI mass for CL-9 at this lower distance is $\sim(7\pm1)\times10^{3}$ $M_{\odot}$, which is similar to the total HI mass of the best-fit model. 

Thus, we conclude that in the unlikely scenario in which CL-9 was as close as $500$ kpc from us, it would still be possible to find a $\Lambda$CDM RELHIC that matches its observed column density. In addition, the maximum allowed distance for CL-9 to contain an HI mass compatible with the system being a RELHIC is $d\sim10$ Mpc, a distance at which CL-9 would still be compatible with a $\Lambda$CDM RELHIC.

\subsubsection{Does CL-9 signal an inner DM deficit?}

\begin{figure}
    \centering
    \includegraphics[]{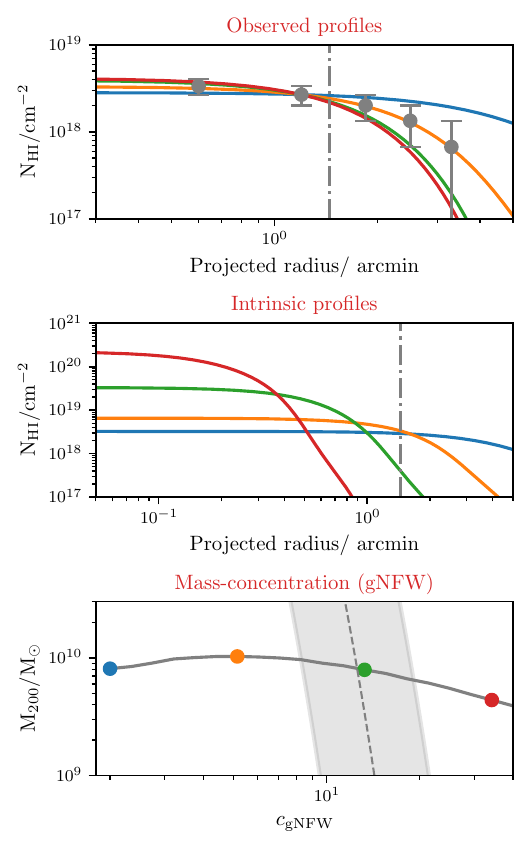}
    \caption{Identical to Fig.~\ref{Fig:LCDM}, but assuming RELHICs are embedded within gNFW halos with $\gamma=0$. See Eq.~\ref{Eq:gNFW}. The solid line in the bottom panel shows the family of models that provide a good fit to the second innermost isocontour.}
    \label{Fig:fits_core}
\end{figure}

An alternative interpretation of the systematic discrepancy between the modeling and observations in the outer regions is that it originates from a deficit of DM relative to a cuspy NFW in the inner regions. To explore this possibility, we now focus on a different model in which the gas in the halo is in hydrostatic equilibrium with a generalized NFW (gNFW) halo, whose logarithmic slope in the inner regions, $\gamma$, is treated as a free parameter:
\begin{equation}
    \rho_{\rm dm}(r)=\rho_{s}\left(\displaystyle\frac{r}{r_{s}}\right)^{-\gamma}\left[1+\left( \displaystyle\frac{r}{r_{s}}\right)\right]^{\gamma-3}
    \label{Eq:gNFW}
\end{equation}
To impose a deficit of DM in the center relative to a cuspy NFW, we enforce a cored inner density profile by setting $\gamma=0$. We consider a grid of RELHIC models, for which we vary both the halo mass and concentration independently of each other. We then fit the models, placed at the same distance of M94 and convolved with the FAST beam, to CL-9's second innermost isocontour. The result of this procedure is shown in Fig.~\ref{Fig:fits_core}. 

With these changes, a RELHIC inhabiting a "cored" DM halo of mass, $M_{200}\sim1.02\times10^{10}$ $M_{\odot}$, and concentration, $c_{\rm gNFW} =r_{200}/r_{s} = 5.15$, matches observations (see the orange line in the top panel of Fig.~\ref{Fig:fits_core}). Other models, found by varying both the halo mass and concentration until they match the central column density, fit the observed profile very poorly. Although the concentration of the best model is off of the~\cite{Ludlow2016} mass-concentration relation (shown in the bottom panel of Fig.~\ref{Fig:fits_core}), there is no reason why a cored profile should follow this relation. 

\begin{figure}
    \centering
    \includegraphics[]{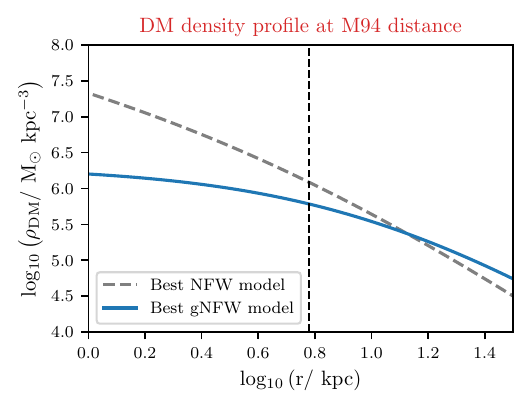}
    \caption{Mean DM density profile of the best-fit $\Lambda$CDM RELHIC model at the fiducial distance of M94 (grey dashed line) compared with the best-fit gNFW RELHIC model. The DM content between the models differs by more than a factor of 2 below $\sim6$ kpc (vertical line).}
    \label{Fig:dm_density}
\end{figure}

The derived halo parameters thus imply a significant DM mass deficit greater than a factor of $2$ for radii, $r\lesssim6$ kpc, compared to $\Lambda$CDM expectations. This is shown in Fig.~\ref{Fig:dm_density}, in which we plot the mean DM density profile of the best $\Lambda$CDM RELHIC and the best gNFW RELHIC. These parameters are uncomfortably large compared with the values expected from self-interacting DM (SIDM) models and difficult to reconcile with $\Lambda$CDM without a bright stellar counterpart. For example,~\cite{Elbert2015} finds that the largest radius at which SIDM halos depart from $\Lambda$CDM is roughly a factor 3 smaller ($\sim2$ kpc). In addition, if CL-9 hosted a stellar counterpart, its low stellar mass would make it difficult for supernova-driven winds to perturb its inner DM at such large distances~\citep[e.g.][]{DiCintio2014, Tollet2016, Robles2017}. Therefore, if CL-9 is confirmed to be a RELHIC and further observations confirm the extended mass deficit, we anticipate challenges reconciling this system not only with $\Lambda$CDM but also with SIDM models. 

\section{Summary and Conclusions}

In this work, we explored whether the recently discovered extended HI gas cloud, CL-9, is consistent with being a $\Lambda$CDM RELHIC. We find that CL-9's properties are consistent with the system being a RELHIC, as recently argued by~\citetalias{Zhou2023}. The match between the model column densities and observations, together with the large projected distance between CL-9 and M94, the round shape of CL-9's isocontours, the lack of a luminous counterpart, the small broadening of the emission line, and the total HI mass make CL-9 the first firm RELHIC candidate in the local Universe. The analysis of this system demonstrates the potential of these objects as cosmological probes.

We also find that CL-9's observations are limited by the FAST beam, whose size is comparable to that of the expected HI core of the most massive RELHICs at the fiducial distance. However, given the high sensitivity of the column density (and total HI mass) to halo mass for RELHICs, we conclude with high confidence that the observed system must inhabit a DM halo with mass in the range $4\times10^9\lesssim M_{200}$/$M_{\odot}$ $\lesssim5\times10^{9}$ if its DM content follows a cuspy NFW profile. This conclusion is based on matching CL-9's total HI mass (and central column density) and, therefore, is independent of whether the system is marginally resolved or unresolved. 

Taken at face value, the marginally resolved  CL-9 column density profile is systematically more extended than expected for a $\Lambda$CDM RELHIC. If confirmed, this may suggest a slightly more massive halo, $M_{200}\sim10^{10}$ $M_{\odot}$, but with a large inner core rather than a cusp. However, before drawing robust conclusions, it is crucial to observe CL-9 with higher spatial resolution.

We envision a series of observations that may help constrain CL-9's parameters and nature. Firstly, the high sensitivity and smaller beam make the MeerKAT radio telescope an obvious choice to constrain CL-9's column density profile better. However, the high declination of CL-9 ($\delta\sim+40^{o}$) places the system at the limit of what can be observed with MeerKAT. CL-9 is also at the reach of the Very Large Array (VLA), an instrument that would increase the spatial resolution of the observed profile. In addition, further observations with FAST may help decrease the beam's impact.

Secondly, the derived halo mass for CL-9 makes the system an excellent candidate to look for the predicted RELHICs' ring-shaped H$\alpha$ emission counterparts~\cite[][]{Sykes2019}. These observations could be performed with narrow-band H$\alpha$ filters on the DragonFly Telephoto Array~\citep{Abraham2014} and would provide data to constrain further the system's DM content and the local intensity of the UVB. 

Thirdly, follow-up observations with the Hubble Space Telescope that go fainter than the limit of the DESI LS may help to elucidate whether CL-9 has a luminous counterpart. 

Finally, observations of bright background sources that intersect CL-9 could help characterize the system's metallicity, thus helping to constrain the likelihood of CL-9 hosting a stellar counterpart.  

There is a high probability that CL-9 contains a luminous galaxy in its center. At the inferred mass, we expect more than $90 \%$ of the halos to host galaxies~\citep[see, e.g.][]{Sawala2016, Benitez-Llambay2017, Benitez-Llambay2020}. Detecting a stellar counterpart would help constrain and break the current degeneracies and assess the quality of our predictions.

Regardless of whether or not CL-9 has a stellar counterpart, its low stellar content, together with its HI reservoir, will still allow us to put joint constraints on its underlying DM distribution. Pursuing this path, although arduous, may be highly rewarding in the end. It will provide a unique opportunity to challenge $\Lambda$CDM and our fundamental understanding of how galaxies form at the smallest scales.
\\
\\
We thank the anonymous referee for a constructive review that helped improve our presentation. ABL acknowledges support from the European Research Council (ERC) under the European Union's Horizon 2020 research and innovation program (GA 101026328). 

\bibliography{references}{}
\bibliographystyle{aasjournal}

\end{document}